\documentclass[3p,times,procedia]{elsarticle}
\usepackage{nupha_ecrc}

\usepackage{enumerate}
\usepackage{amsmath}
\usepackage{amssymb}

\usepackage[utf8]{inputenc}
\usepackage[T1]{fontenc}

\volume{00}

\firstpage{1}

\journalname{Nuclear Physics A}

\runauth{}

\jid{nupha}

\jnltitlelogo{Nuclear Physics A}

\usepackage{amssymb}

\usepackage[figuresright]{rotating}

\newcommand{\MeV}{\textrm{MeV}}

\begin{document}

\begin{frontmatter}


\title{Equation of state for QCD with a critical point from the 3D Ising Model \tnoteref{}}
\author{Paolo Parotto}
\ead{pparotto@uh.edu}
\address{Department of Physics, University of Houston, Houston, TX 77204, USA \vspace{-0cm} \fnref{}}

\dochead{XXVIIth International Conference on Ultrarelativistic Nucleus-Nucleus Collisions\\ (Quark Matter 2018)}





\begin{abstract}
Current knowledge of the finite-density QCD equation of state from first principles is limited to a Taylor expansion in the baryonic chemical potential around $\mu_B=0$. By means of a scaling form for the equation of state of the 3D Ising model and a non-universal, parametrized map to QCD coordinates, we construct a family of equations of state matching state of the art first principle Lattice QCD calculations and including the correct critical behavior, which can be readily employed in hydrodynamical simulations of heavy ion collisions at finite density, covering most of the BES range at RHIC. This contribution reports on work done within the Fluctuations/Equation of State working group of the BEST Collaboration.
\end{abstract}

\end{frontmatter}


\section{Introduction}
\label{sec:intro}
The determination of the phase structure of QCD is nowadays one of the most important goals of high energy nuclear physics research, stimulating strong efforts from both the theoretical and the experimental communities. In recent years, Lattice QCD calculations have provided increasingly accurate quantitative results for the thermodynamics of strongly interacting matter, for baryon-antibaryon symmetric matter, establishing the presence of a continuous, smooth phase transition between confined hadronic matter and deconfined quark-gluon plasma at a temperature of $T \simeq 155 \, \MeV$. It is believed that such transition would become of the first order at higher baryonic densities, implying the presence of a critical point; past works have shown that such a critical point would belong to the same universality class as the three dimensional Ising model \cite{Rajagopal:1992qz}.

Because of the sign problem of Lattice QCD at non-zero chemical potential, it is not possible to approach the search for the QCD critical point directly from first principles. Current knowledge of the QCD equation of state (EoS) from principles at finite chemical potential is limited to a Taylor expansion around $\mu_B = 0$.

On the experimental side, this search has recently experienced a peak in productivity in view of the BES-II program, which will take place at the Relativistic Heavy Ion Collider in the next couple years, and has as its main intent to locate the critical point.

From a theoretical point of view, the study of heavy-ion collisions and the interpretation of experimental data is majorly  performed via hydrodynamic simulations, which take as their main ingredient the equation of state of QCD matter, driving the evolution of the system. The study of the presence of the critical point in the phase diagram, in particular in the density range accessible to the BES-II program, cannot prescind from an equation of state including the correct critical behavior.

\section{The procedure}
The purpose of the work exposed here (see \cite{Parotto:2018pwx} for the complete discussion, and references therein) is to generate a family of equations of state in a parametric form, which exactly match the known Lattice QCD results at vanishing baryon density, and contain critical behavior in the right universality class. 

The procedure can be summarized as follows:
\begin{enumerate}[i)]
\item \textit{Implement the scaling behavior of the 3D Ising model EoS:}

In the vicinity of the critical point, one can use the following parametrization for the magnetization $M$, the magnetic field $h$ and the reduced temperature $r=\left( T- T_C \right)/T_C$ \cite{Guida:1996ep,Nonaka:2004pg}:
\begin{align}
M &= M_0 R^\beta \theta \, \, , & \quad h &= h_0 R^{\beta \delta} \tilde{h}(\theta) \, \, , & \quad
r &= R (1- \theta^2) \, \, , 
\end{align}
where $M_0$, $h_0$ are normalization constants, $\tilde{h}(\theta) = \theta (1 + a \theta^2 + b \theta^4)$  with $a=-0.76201$, $b=0.00804$, $\beta$ and $\delta$ are 3D Ising critical exponents, and the parameters take on the values $R \geq 0$, $\left| \theta \right| \leq \theta_0 \simeq 1.154$, $\theta_0$ being the first non-trivial zero of $\tilde{h}(\theta)$. 

\item \textit{Define a non-universal map from the 3D Ising model phase diagram to the QCD one:}

In order to transfer the critical thermodynamics to QCD, a non-universal mapping is needed between Ising variables $(h,r)$ and QCD coordinates $(T,\mu_B)$. The most general linear transformation allowing this makes use of six parameters:
\begin{equation}
\begin{aligned}
\frac{T - T_C}{T_C} &=  w \left( r \rho \,  \sin \alpha_1  + h \, \sin \alpha_2 \right) \, \, ,  \\
\frac{\mu_B - \mu_{BC}}{T_C} &=  w \left( - r \rho \, \cos \alpha_1 - h \, \cos \alpha_2 \right) \, \, .
\end{aligned} 
\end{equation}
where $(T_C,\mu_{BC})$ give the chosen location of the critical point, $\alpha_1$ and $\alpha_2$ indicate the relative angle between the $r$ and $h$ axes and the lines of $T= \textit{const.}$, and the parameters $w$ and $\rho$ correspond to a global and relative rescaling of $r$ and $h$. It is possible to reduce the number of parameters, assuming the shape of the chiral transition line to be a parabola (a good approximation in the BES range); the curvature of the transition line and the transition temperature at $\mu_B=0$ calculated from Lattice QCD can be used \cite{Bellwied:2015rza}. In this work, we will show, as an example, results for the following choice of parameters:
\begin{equation}\label{eq:params}
\begin{aligned}
\mu_{BC} &= 350 \, \MeV \, \, , \quad & T_C &\simeq 143.2 \, \MeV \, \, , \quad & \alpha_1 &\simeq 3.85 ^\circ \, \, ,  \\ 
\alpha_2 &\simeq 93.85 ^\circ \, \, , \quad & w &= 1 \, \, , \quad & \rho &= 2 \, \, .
\end{aligned}
\end{equation}

\item \textit{Estimate the contribution to Taylor coefficients from the 3D Ising model critical point:} 

We decompose the Taylor coefficients from Lattice QCD as the sum of an ``Ising'' contribution from the critical point placed onto the phase diagram, and a ``Non-Ising'' one:
\begin{equation}\label{eq:coeffs}
T^4 \, c_n^{LAT}(T) = T^4 \, c_n^{\textrm{Non-Ising}}(T) + T_C^4 \, c_n^{\textrm{Ising}}(T)
\end{equation}
where the $c_n^{LAT}(T)$ are a parametrization of continuum extrapolated Lattice QCD results \cite{Borsanyi:2010cj,Bellwied:2015lba}, continued at low $T$ with Hadron Resonance Gas model calculations using an up to date hadronic spectrum \cite{Alba:2017mqu}.


\item \textit{Reconstruct the full pressure:}

Once the ``Non-Ising'' coefficients are calculated using Eq. (\ref{eq:coeffs}), the full pressure can be reconstructed over the whole phase diagram with:
\begin{equation}\label{eq:fullP}
P(T,\mu_B) = T^4 \sum_n c_n^{\text{Non-Ising}}  \left( T \right) \left( \frac{\mu_B}{T} \right)^n + T_C^4 P^{\text{Ising}} \left( T,\mu_B \right) \, \, .
\end{equation}

From the full pressure in Eq. (\ref{eq:fullP}), all thermodynamic quantities of interest can be calculated. In Fig. \ref{fig:PressEntrFinal} we show the full pressure, as well as the speed of sound calculated with our procedure with the parameter choice in Eq. (\ref{eq:params}).

\begin{figure}[h]
\center
\includegraphics[width=.42\textwidth]{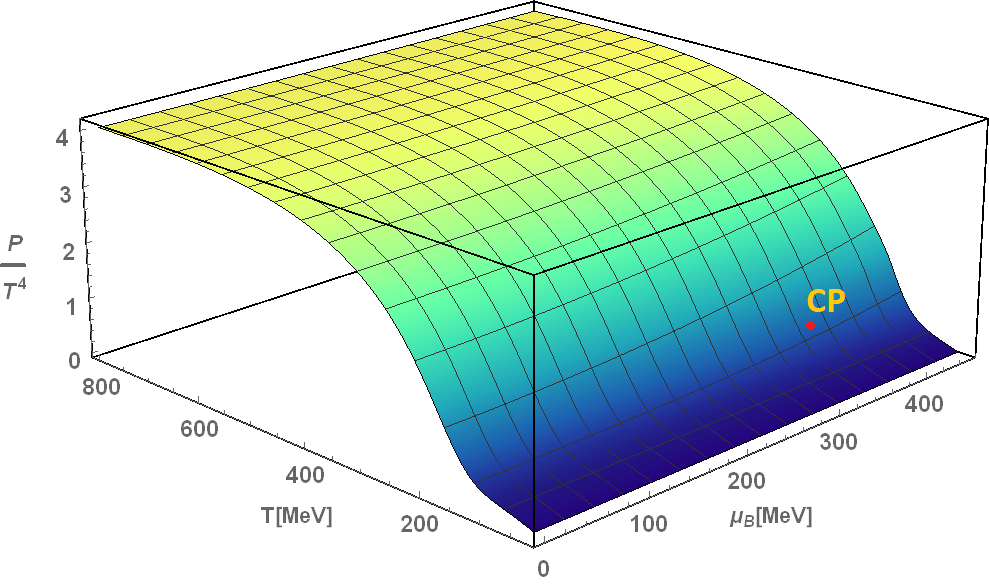} \hspace{1cm}
\includegraphics[width=.42\textwidth]{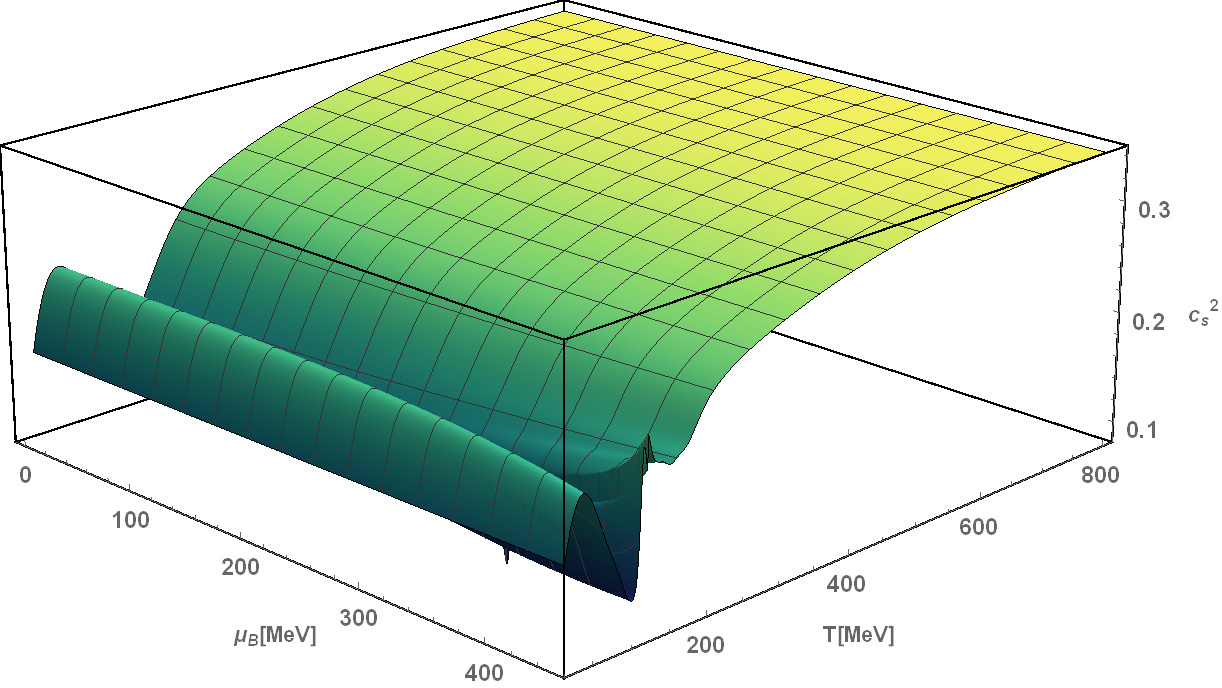}
\caption{Full pressure and speed of sound for the choice of parameters in Eq. (\ref{eq:params}).}
\label{fig:PressEntrFinal}
\end{figure}

\end{enumerate}

The procedure summarized in this contribution (see \cite{Parotto:2018pwx} for the complete discussion) allowed us to construct a family of equations of state for QCD matching Lattice QCD results at vanishing chemical potential, and containing a critical point in the correct universality class in a parametrized fashion; our results can be readily utilized as input in hydrodynamic simulations of heavy ion collisions.


\section*{Acknowledgements}
This material is based upon work supported by the National Science Foundation under Grants No. PHY-1654219 and OAC-1531814 and by the U.S. Department of Energy, Office of Science, Office of Nuclear Physics, within the framework of the Beam Energy Scan Theory (BEST) Topical Collaboration. An award of computer time was provided by the INCITE program. This research used resources of the Argonne Leadership Computing Facility, which is a DOE Office of Science User Facility supported under Contract No. DE-AC02-06CH11357. The authors gratefully acknowledge the use of the Maxwell Cluster and the advanced support from the Center of Advanced Computing and Data Systems at the University of Houston.





\bibliographystyle{elsarticle-num}
\bibliography{references}

\begin{thebibliography}{1}
\expandafter\ifx\csname url\endcsname\relax
  \def\url#1{\texttt{#1}}\fi
\expandafter\ifx\csname urlprefix\endcsname\relax\def\urlprefix{URL }\fi
\expandafter\ifx\csname href\endcsname\relax
  \def\href#1#2{#2} \def\path#1{#1}\fi

\bibitem{Rajagopal:1992qz}
K.~Rajagopal, F.~Wilczek, {Static and dynamic critical phenomena at a second
  order QCD phase transition}, Nucl. Phys. B399 (1993) 395--425.
\newblock \href {http://arxiv.org/abs/hep-ph/9210253}
  {\path{arXiv:hep-ph/9210253}}, \href
  {http://dx.doi.org/10.1016/0550-3213(93)90502-G}
  {\path{doi:10.1016/0550-3213(93)90502-G}}.

\bibitem{Parotto:2018pwx}
P.~Parotto, M.~Bluhm, D.~Mroczek, M.~Nahrgang, J.~Noronha-Hostler,
  K.~Rajagopal, C.~Ratti, T.~Schäfer, M.~Stephanov, {Lattice-QCD-based
  equation of state with a critical point}\href
  {http://arxiv.org/abs/1805.05249} {\path{arXiv:1805.05249}}.

\bibitem{Guida:1996ep}
R.~Guida, J.~Zinn-Justin, {3-D Ising model: The Scaling equation of state},
  Nucl. Phys. B489 (1997) 626--652.
\newblock \href {http://arxiv.org/abs/hep-th/9610223}
  {\path{arXiv:hep-th/9610223}}, \href
  {http://dx.doi.org/10.1016/S0550-3213(96)00704-3}
  {\path{doi:10.1016/S0550-3213(96)00704-3}}.

\bibitem{Nonaka:2004pg}
C.~Nonaka, M.~Asakawa, {Hydrodynamical evolution near the QCD critical end
  point}, Phys. Rev. C71 (2005) 044904.
\newblock \href {http://arxiv.org/abs/nucl-th/0410078}
  {\path{arXiv:nucl-th/0410078}}, \href
  {http://dx.doi.org/10.1103/PhysRevC.71.044904}
  {\path{doi:10.1103/PhysRevC.71.044904}}.

\bibitem{Bellwied:2015rza}
R.~Bellwied, S.~Borsanyi, Z.~Fodor, J.~Guenther, S.~D. Katz, C.~Ratti, K.~K.
  Szabo, {The QCD phase diagram from analytic continuation}, Phys. Lett. B751
  (2015) 559--564.
\newblock \href {http://arxiv.org/abs/1507.07510} {\path{arXiv:1507.07510}},
  \href {http://dx.doi.org/10.1016/j.physletb.2015.11.011}
  {\path{doi:10.1016/j.physletb.2015.11.011}}.

\bibitem{Borsanyi:2010cj}
S.~Borsanyi, G.~Endrodi, Z.~Fodor, A.~Jakovac, S.~D. Katz, S.~Krieg, C.~Ratti,
  K.~K. Szabo, {The QCD equation of state with dynamical quarks}, JHEP 11
  (2010) 077.
\newblock \href {http://arxiv.org/abs/1007.2580} {\path{arXiv:1007.2580}},
  \href {http://dx.doi.org/10.1007/JHEP11(2010)077}
  {\path{doi:10.1007/JHEP11(2010)077}}.

\bibitem{Bellwied:2015lba}
R.~Bellwied, S.~Borsanyi, Z.~Fodor, S.~D. Katz, A.~Pasztor, C.~Ratti, K.~K.
  Szabo, {Fluctuations and correlations in high temperature QCD}, Phys. Rev.
  D92~(11) (2015) 114505.
\newblock \href {http://arxiv.org/abs/1507.04627} {\path{arXiv:1507.04627}},
  \href {http://dx.doi.org/10.1103/PhysRevD.92.114505}
  {\path{doi:10.1103/PhysRevD.92.114505}}.

\bibitem{Alba:2017mqu}
P.~Alba, et~al., {Constraining the hadronic spectrum through QCD thermodynamics
  on the lattice}, Phys. Rev. D96~(3) (2017) 034517.
\newblock \href {http://arxiv.org/abs/1702.01113} {\path{arXiv:1702.01113}},
  \href {http://dx.doi.org/10.1103/PhysRevD.96.034517}
  {\path{doi:10.1103/PhysRevD.96.034517}}.

\end{thebibliography}







\end{document}